# A Comprehensive Review on Non-Neural Networks Collaborative Filtering Recommendation Systems

**Carmel Wenga**[a,b], **Majirus Fansi**[b], **Sébastien Chabrier**[a], **Jean-Martial Mari**[a], **Alban Gabillon**[a].

[a] *Université de la Polynésie Française, Tahiti, French Polynesia - {alban.gabillon, sebastien.chabrier, jean-martial.mari}@upf.pf*

[b] *NzhinuSoft - {carmel.wenga, majirus.fansi}@nzhinusoft.com*

## Abstract

Over the past two decades, recommender systems have attracted a lot of interest due to the explosion in the amount of data in online applications. A particular attention has been paid to collaborative filtering, which is the most widely used in applications that involve information recommendations. Collaborative filtering (CF) uses the known preference of a group of users to make predictions and recommendations about the unknown preferences of other users (recommendations are made based on the past behavior of users). First introduced in the 1990s, a wide variety of increasingly successful models have been proposed. Due to the success of machine learning techniques in many areas, there has been a growing emphasis on the application of such algorithms in recommendation systems. In this article, we present an overview of the CF approaches for recommender systems, their two main categories, and their evaluation metrics. We focus on the application of classical Machine Learning algorithms to CF recommender systems by presenting their evolution from their first use-cases to advanced Machine Learning models. We attempt to provide a comprehensive and comparative overview of CF systems (with python implementations) that can serve as a guideline for research and practice in this area.

## 1 Introduction

In everyday life, people often refer to the opinions and advice of others when it comes to purchasing for a service from a company, buying an item in a marketplace or a store, watching a movie at a cinema, etc. In general, they are influenced by the appreciation of their neighborhood before purchasing (or not) an item. In some cases, they may explicitly receive recommendations from others on what to buy or not. This natural and social process is a recommendation one which implies that the different actors collaborate in order to filter their interests according to the experience of others. It is a collaborative filtering process, where people collaborate to help one another perform filtering by recording their appreciations on items they purchased. In 1992, Goldberg *et al.* introduced the Collaborative Filtering algorithm to help people filtering their emails or documents according to the evaluations of others to these documents [Goldberg,Nichols,Oki & Terry 1992]. Today, recommender systems implement and amplify this simple process and apply it to many areas such as e-commerce web-site (Amazon), social networks (Facebook), streaming applications (Netflix), etc.

In 2000, K. Goldberg *et al.* defined the term *Collaborative Filtering* (CF) as "techniques that use the known preferences of a group of users to predict the unknown preferences of a new user" [Goldberg,Roeder,Gupta & Perkins 2001]. According to [Breese,Heckerman & Kadie 1998], CF recommender systems are built on the assumption that a good way to find interesting content is to find other people who have similar interests, and then recommend titles that those similar users like. The fundamental concept of CF can then by defined as : *if users $u$ and $v$ rate $n$ items similarly or have similar behaviors, they share similar tastes and will hence rate other items similarly* [Goldberg et al. 2001]. User's preferences can be expressed through *ratings, likes, clicks,* etc. CF algorithms hence use databases of such preferences records made by users on different items. Typically, preferences of a list of users on a list of items are converted into a *user-item preferences matrix*. Table 1 presents an example of user-item preferences matrix $R$, where preferences are ratings scores made by users on items and each cell $R_{u,i}$ denotes the rating score of user $u$ on item $i$. CF algorithms use such user-item observed interactions to predict the unobserved ones [Vucetic & Obradovic 2005].

*Table 1: An example of user-item rating matrix $R$ of dimension $m \times n$, where $m$ is the number of users and $n$ the number of items. Each row represents all ratings of a given user and each column ratings given to a particular item. $R_{u,i}$ is the rating score of user $u$ on item $i$.*

|       | $i_1$ | $i_2$ | $\cdots$ | $i_j$ | $\cdots$ | $i_n$ |
|-------|-------|-------|----------|-------|----------|-------|
| $u_1$ | 5     | 3     | 1        | 2     |          | 1     |
| $u_2$ |       | 2     |          |       |          | 4     |
| $\vdots$ |    |       | 1        |       |          |       |
| $u_i$ | 3     | 4     | 1        | 2     |          | ?     |
| $\vdots$ |    |       |          |       |          |       |
| $u_m$ |       |       | 2        |       |          | 5     |

There are two majors' types of CF recommender systems : Memory-based CF (Section 2) and Model-based CF (Section 3). Memory-based CF operates over the entire database of user's preferences and performs computations on it each time a new prediction is needed [Breese et al. 1998; Vucetic & Obradovic 2005]. The most common representative algorithms for Memory based CF are neighborhood-based algorithms. Those algorithms compute similarities between users in order to predict what to recommend. A typical neighborhood-based CF is the *GroupLens* system [Resnick,Iacovou,Suchak,Bergstrom & Riedl 1994] in which numerical ratings assigned by users on articles are correlated in order to determine which users' ratings are most similar to each other. Based on these similarities, *GroupLens* predicts how well users will like new articles. Memory-based CF can produce quite good predictions and recommendations, provided that a suitable function is used to measure similarities between users [Hernando,Bobadilla & Ortega 2016]. According to Ansari et al. [Ansari,Essegaier & Kohli 2000], such algorithms can also increase customer loyalty, sales, advertising revenues, and the benefit of targeted promotions. However, an important drawback of Memory-based CF is that they do not use scalable algorithms [Resnick et al. 1994; Hernando et al. 2016]. Therefore, they would have high latency in giving predictions for active users in a system with a large number of requests that should be processed in real-time [Vucetic & Obradovic 2005] .

While Memory-based algorithms perform computations over the entire database each time a new prediction is needed, Model-based algorithms (Section 3) in contrast, use the user-item interactions to estimate or learn a model, which is used for predictions and recommendations [Breese et al. 1998; Parhi,Pal & Aggarwal 2017]. Most common Model-based approaches include Dimensionality Reduction models [Billsus & Pazzani 1998; Sarwar,Karypis,Konstan & Riedl 2000; Koren,Bell & Volinsky 2009; Hernando et al. 2016], Clustering models [Ungar & Foster 1998; Chee,Han & Wang 2001], Bayesian models [Breese et al. 1998] or Regression models [Mild & Natter 2002; Kunegis & Albayrak 2007]. Most recent Model-based CF are based on Neural Networks and Deep Learning algorithms [Sedhain,Menon,Sanner & Xie 2015; He,Liao,Zhang,Nie,Hu & Chua 2017a; Rawat & Wang 2017; He,Du,Wang,Tian,Tang & Chua 2018; Mu 2018; Zhang,Luo,Zhang & Wu 2018; Yu,Si,Hu & Zhang 2019]. However, in this review, we do not present CF recommender systems based on Neural Networks and Deep Learning approaches which deserve complete separate study.

One of the major challenges of CF algorithms is the *cold start* problem, which is the lack of information necessary to produce recommendations for new users or new items [Su & Khoshgoftaar 2009]. Beside collaborative filtering, content-based filtering [Pazzani & Billsus 2007] is another powerful recommendation approach. While CF recommender systems use interactions between users and items, content-based filtering uses users' profiles (gender, age, profession, ...) or items' contents (title, description, images, ...) to produce recommendations. Combining collaborative filtering to content-based filtering in a hybrid model helps to address the cold start problem and thereby improve recommendation performance [Su & Khoshgoftaar 2009].

**Previous works**

In [Su & Khoshgoftaar 2009], the authors present an overview of collaborative filtering algorithms built before 2009. They present the basics of CF algorithms as well as their two main techniques. However, [Su & Khoshgoftaar 2009] do not present advanced machine learning techniques such as Matrix Factorization models [Mnih & Salakhutdinov 2008; Koren et al. 2009; Lakshminarayanan,Bouchard & Archambeau 2011; Hernando et al. 2016; Abdollahi & Nasraoui 2017] which are highly used in today's recommendation systems. In [Herlocker,Konstan,Terveen & Riedl 2004; Mnih & Salakhutdinov 2008; Ekstrand,Riedl & Konstan 2011; Bobadilla,Ortega,Hernando & Gutiérrez 2013], the authors present a few classical dimensionality reduction models applied to recommendation systems. However, more complex dimensionality reduction models have been proposed, such as Non-negative Matrix Factorization [Gillis 2014; Hernando et al. 2016] and Explainable Matrix Factorization [Abdollahi & Nasraoui 2017; Wang,Tian,Zhu & Wu 2018].

In this article, we follow the methodology of [Su & Khoshgoftaar 2009; Ekstrand et al. 2011] and present a comprehensive review of collaborative filtering recommender systems based on classical Machine Learning techniques (non-Neural Networks models). This review relates the evolution of CF algorithms from their first use cases (basic CF algorithms) to advanced machine learning techniques. We provide an overview of CF techniques and the most important evaluation metrics to be considered when building a recommendation system. We propose a Python/Numpy/Pandas implementation (available on github[1]) for each model presented in this article, as well as the comparison of their performances on the Movie Lens[2] benchmark datasets.

The rest of this document is structured as follows : Section 2 and Section 3 respectively present Memory and Model-based CF; Section 4 presents evaluation metrics of recommender systems; In Section 5, we present the implementation details of our comparative experimentation and we evaluate and compare the models performances. Section 6 concludes this study.

---

1   Numpy/Pandas implementation of models presented in this paper (https://github.com/nzhinusoftcm/review-on-collaborative-filtering)
2   **MovieLens** (https://grouplens.org/datasets/movielens/) is a benchmark dataset collected and made available in several sizes (100k, 1M, 25M) by the GroupLens Research team.

# 2 Memory-based Collaborative Filtering

Memory-based algorithms compute predictions using the entire user-item interactions directly on the basis of similarity measures. They can be divided into two categories, User-based CF [Herlocker,Konstan,Borchers & Riedl 1999] and Item-based CF [Sarwar,Karypis,Konstan & Reidl 2001; Su & Khoshgoftaar 2009; Desrosiers & Karypis 2011]. In User-based model, a user $u$ is identified by his ratings $R_u$ on the overall items while in Item-based model, an item i is identified by all the ratings $r_i$ attributed by users on that item. For instance, in Table 2, $R_{u_1} = \{1, 5, 0, 2, 4, 0\}$ and $R_{i_4} = \{2, 5, 0, 5\}$; where 0 represents non-observed values. More generally, with $m$ users and $n$ items, users and items ratings are respectively $\mathbb{R}^n$ and $\mathbb{R}^m$ vectors. Table 2 presents an overview of memory-based CF.

*Table 2: A sample matrix of ratings*

|       | $i_1$ | $i_2$ | $i_3$ | $i_4$ | $i_5$ | $i_6$ |
|-------|-------|-------|-------|-------|-------|-------|
| $u_1$ | 1     | 5     |       | 2     | 4     | ?     |
| $u_2$ | 4     | 2     |       | 5     | 1     | 2     |
| $u_3$ | 2     | 4     | 3     |       |       | 5     |
| $u_4$ | 2     | 4     |       | 5     | 1     | ?     |

Memory-based recommendation systems (both User-based and Item-based collaborative filtering) usually proceeds in three stages : Similarity computations, Ratings predictions and Top-N recommendations.

## 2.1 Similarity Computation

In either User-based or Item-based CF, users and items are modeled with vector-spaces. In User-based CF (resp. Item-based CF), users are treated as vectors of $n$-dimensional items (resp. $m$-dimensional users) space. The similarity between two users (or two items) is known by computing the similarity between their corresponding vectors. For User-based CF, the idea is to compute the similarity $w_{u,v}$ between users $u$ and $v$ who have both rated the same items. In contrast, Item-based CF performs similarity $w_{i,j}$ between items $i$ and $j$ which have both been rated by the same users [Sarwar et al. 2001; Su & Khoshgoftaar 2009]. Similarity computation methods include Correlation-based similarity such as *Pearson correlation* [Resnick et al. 1994], *Cosine-based similarity* [Dillon 1983], *Adjusted Cosine Similarity* [Sarwar et al. 2001]. This section presents the most widely used similarity metrics for CF algorithms.

### 2.1.1 Pearson Correlation similarity

The correlation between users $u$ and $v$ (or between items $i$ and $j$) is a weighted value that ranges between -1 and 1 and indicates how much user $u$ would agree with each of the ratings of user $v$ [Resnick et al. 1994]. Correlations are mostly performed with the *Pearson correlation* metric (equation 1). The idea here is to measure the extent to which there is a linear relationship between the two users $u$ and $v$. It's the covariance (the joint variability) of $u$ and $v$.

$$w_{u,v} = \frac{\sum_{i \in I}(R_{u,i} - \bar{R}_u)(R_{v,i} - \bar{R}_v)}{\sqrt{\sum_{i \in I}(R_{u,i} - \bar{R}_u)^2}\sqrt{\sum_{i \in I}(R_{v,i} - \bar{R}_v)^2}}, \qquad (1)$$

where $I$ is the set of items that have both been rated by users $u$ and $v$, $R_{u,i}$ the rating score of user $u$ on item $i$ and $\bar{R}_u$ is the average rating of co-rated items of user $u$. For example, according to Table 2, $w_{u_1,u_2} = -0.8$, $w_{u_1,u_3} = 1$ and $w_{u_1,u_4} = 0$. This means that, user $u_1$ tends to disagree with $u_2$, agree with $u_3$ and that his ratings are not correlated at all with those of $u_4$.

For the Item-based CF, all summations and averages are computed over the set $U$ of users who both rated items $i$ and $j$ [Sarwar et al. 2001]. The correlations similarity for Item-based CF is then given by equation 2.

$$w_{i,j} = \frac{\sum_{u \in U}(R_{u,i} - \bar{R}_i)(R_{u,j} - \bar{R}_j)}{\sqrt{\sum_{u \in U}(R_{u,i} - \bar{R}_i)^2}\sqrt{\sum_{u \in U}(R_{u,j} - \bar{R}_j)^2}}. \qquad (2)$$

Correlation based algorithms work well when data are not highly sparse. However, when having a huge number of users and items, the user-item interaction matrix is extremely sparse. This makes predictions based on the *Pearson correlation* less accurate. In [Su,Khoshgoftaar,Zhu & Greiner 2008], the authors proposed an *imputation-boosted CF* (*IBCF*) which first predicts and fills-in missing data. The *Pearson correlation-based CF* is then used on that completed data to predict a user rating on a specific item.

## 2.1.2 Cosine-based similarity

The cosine similarity between users $u$ and $v$ (modeled by their ratings vectors $\vec{R}_u$ and $\vec{R}_v$) is measured by computing the cosine of the angle between these two vectors [Sarwar et al. 2001]. Formally, if $R$ is the $m \times n$ user-item ratings matrix, then the similarity between users $u$ and $v$, denoted by $w_{u,v}$ is computed by equation 3 :

$$w_{u,v} = \frac{\vec{R}_u \cdot \vec{R}_v}{\|\vec{R}_u\|_2 * \|\vec{R}_v\|_2} = \frac{\sum_{i \in I} R_{u,i} R_{v,i}}{\sqrt{\sum_{i \in I}(R_{u,i})^2}\sqrt{\sum_{i \in I}(R_{v,i})^2}}, \qquad (3)$$

where $R_{u,i}$ denotes the rating of user $u$ on item $i$.

The basic cosine measure given previously can easily be adapted for item's similarities. However, when applied to Item-based cases, it has an important drawback due to the fact that the differences in ratings scale between different users are not taken into account [Sarwar et al. 2001]. Indeed, some users would tend to give higher ratings than others, making their ratings scales to be different [Sarwar et al. 2001; Su & Khoshgoftaar 2009]. The *Adjusted Cosine Similarity* addresses this drawback by subtracting the corresponding user's mean rating $\bar{r}_u$ from each co-rated pair [Sarwar et al. 2001] . Formally, the similarity between items $i$ and $j$ using the *Adjusted Cosine Similarity* is given by equation 4.

$$w_{i,j} = \frac{\sum_{u \in U}(R_{u,i} - \bar{R}_u)(R_{u,j} - \bar{R}_u)}{\sqrt{\sum_{u \in U}(R_{u,i} - \bar{R}_u)^2}\sqrt{\sum_{u \in U}(R_{u,j} - \bar{R}_u)^2}}. \qquad (4)$$

*Conditional Probability* [Karypis 2001; Deshpande & Karypis 2004] is an alternative way of computing the similarity between pairs of items $i$ and $j$. The conditional probability of $i$ given $j$ denoted by $P(i|j)$, is the probability of purchasing item $i$ given that item $j$ has already been purchased [Karypis 2001]. The idea is to use $P(i|j)$ when computing $w_{i,j}$. However, since the conditional probability is not symmetric ($P(i|j)$ may not be equivalent to $P(j|i)$), $w_{i,j}$ and $w_{j,i}$ will not be equal either. So, a particular attention must be taken that this similarity metric is used in the correct orientation [Karypis 2001; Ekstrand et al. 2011]. Among all these similarity metrics, *Adjusted Cosine* showed to be better (in terms of Mean Absolute Error) than pure *Cosine* and *Pearson Correlation* [Sarwar et al. 2001].

## 2.2 Prediction and Recommendation

Memory-based algorithms provide two types of recommendations pipelines : *Predictions* and *Top-N recommendations* [Sarwar et al. 2001]. Predictions and recommendation are the most important steps in collaborative filtering algorithms [Sarwar et al. 2001]. They are generated from the average of the nearest users' ratings, weighted by their corresponding correlations with the referenced user [Su & Khoshgoftaar 2009].

### 2.2.1 Prediction with weighted sum

Predictions for an active user $u$, on a certain item $i$ is obtained by computing the weighted average of all the ratings of his nearest users on the target items [Resnick et al. 1994; Su & Khoshgoftaar 2009]. The goal is to determine how much user $u$ will like item $i$. Let us denote by $\hat{R}_{u,i}$ the predicted rating that user $u$ would have given to item $i$ according to the recommender system. $\hat{R}_{u,i}$ is computed using equation 5,

$$\hat{R}_{u,i} = \bar{R}_u + \frac{\sum_{v \in U}(R_{v,i} - \bar{R}_v) \cdot w_{u,v}}{\sum_{v \in U}|w_{u,v}|}, \tag{5}$$

where $\bar{R}_u$ and $\bar{R}_v$ are the average ratings of users $u$ and $v$ on all the other items, and $w_{u,v}$ is the weight or the similarity between users $u$ and $v$. Summation is made over all the users who have rated item $i$. As example, using the user-item ratings of Table 2, let us predict the score for user $u_1$ on item $i_6$ with equation 6.

$$\hat{R}_{u_1,i_6} = \bar{R}_{u_1} + \frac{\sum_{k \in \{2,3\}}(R_{u_k,6} - \bar{R}_{u_k}) \cdot w_{u_1,u_k}}{\sum_{k \in \{2,3\}}|w_{u_1,u_k}|} = 3 + \frac{0.8 + 2}{|-0.8| + |1|} \approx 4.56. \tag{6}$$

This predicted value is reasonable for $u_1$, since he tends to agree with $u_3$ and disagree with $u_2$ (§ 2.1.1). So, the predicted value is closer to 5 (rating of $u_3$ on $i_6$) than 2 (rating of $u_2$ on $i_6$). For Item-based CF, the prediction step uses the *simple weighted average* to predict ratings. The idea is to identify how the active user rated items similar to $i$ and then make a prediction on item $i$ by computing the sum of the ratings (weighted by their corresponding similarities) given by the user on those similar items [Sarwar et al. 2001]. For Item-based CF, $R_{u,i}$ is computed by equation 7:

$$\hat{R}_{u,i} = \frac{\sum_{j \in S^{(i)}} R_{u,j} \cdot w_{i,j}}{\sum_{j \in S^{(i)}}|w_{i,j}|}, \tag{7}$$

where $S^{(i)}$ is the set of items similar to item $i$, $w_{i,j}$ is the correlation between items i and j and $R_{u,j}$ is the rating of user $u$ on item $j$.

### 2.2.2 Top-N Recommendation

The Top-N recommendation algorithm aims to return the list of the most relevant items for a particular user. To recommend items that will be of interest to a given user, the algorithm determines the Top-N items that the user would like. The computation of these Top-N recommendations depends on whether the recommender system is a User-based CF or an Item-based CF.

**User-based Top-N recommendation**

In User-based CF, the set $U$ of all users is divided into many groups $U_G$ in which users in the same group behave similarly. When computing the Top-N items to recommend to an active user $u$, the algorithm first identify the group $U_G$ to which the user $u$ belongs to ($k$ most similar users of $u$). Then it aggregates the ratings of all users in $U_G$ to identify the set $C$ of items purchased by the group as well as the frequency. User-based CF techniques then recommends the $N$ most relevant items in $C$ according to user $u$, and that $u$ has not already purchased [Karypis 2001; Sarwar et al. 2001; Ekstrand et al. 2011].

Despite their popularity, User-based recommendation systems have a number of limitations related to data sparsity, scalability and real-time performance [Karypis 2001; Sarwar et al. 2001; Ekstrand et al. 2011]. Indeed, the computational complexity of these methods grows linearly with the number of users and items which can grow up to several millions in typical e-commerce websites. In such systems, even active users typically interact with under 1% of the items [Sarwar et al. 2001], leading to a high data-sparsity problem. Furthermore, as a user rates and re-rates items, their rating vector will change, along with their similarity to other users. This makes it difficult to find similar users in advance. For this reason, most User-based CF find neighborhoods at run time [Ekstrand et al. 2011] , hence the scalability concern.

**Item-based Top-N recommendation**

Item-based CF, also known as item-to-item CF algorithms have been developed to address the scalability concerns of User-based CF [Karypis 2001; Sarwar et al. 2001; Linden,Smith & York 2003; Ekstrand et al. 2011]. They analyze the user-item matrix to identify relations between the different items, and then use these relations to compute the list of top-N recommendations. The key idea of this strategy is that a customer is more likely to buy items that are similar or related to items he/she has already purchased. [Sarwar et al. 2001]. Since this doesn't need to identify the neighborhood of similar customers when a recommendation is requested, they lead to much faster recommendation engines.

Item-based CF first compute the $k$ most similar items for each item and the corresponding similarities are recorded. The Top-N recommendation algorithm for an active user $u$ that has purchased a set $I_u$ of items is defined by the following steps : (a) Identify the set $C$ of candidate items by taking the union of the $k$ most similar items of $I_u$ and removing items already purchased by user $u$ (items in the set $I_u$). (b) Compute the similarities between each item $c \in C$ and the set $I_u$ as the sum of similarities between all items in $I_u$ and $c$, using only the $k$ most similar item of $i$, $\forall i \in I_u$. (c) Sort the remaining items in C in decreasing order of similarity, to obtain the Item-based Top-N recommendation list [Karypis 2001; Sarwar et al. 2001; Linden et al. 2003; Su & Khoshgoftaar 2009].

Item-based algorithms are suitable to make recommendations to users who have purchased a set of items. For example, this algorithm can be applied to make top-N recommendations for items in a user shopping cart [Linden et al. 2003]. The effectiveness of Item-based CF depends on the method used to compute similarity between various items. In general, the similarity between two items $i$ and $j$ should be high if there are a lot of users that have purchased both of them, and it should be low if there are few such users [Linden et al. 2003; Deshpande & Karypis 2004]. Some other criteria to consider when computing similarity between items are specified in [Deshpande & Karypis 2004].

The key advantages of Item-based over User-based collaborative filtering are scalability and performance. The algorithm's online component scales independently of the catalog size or of the total number of users. It depends only on how many items the user has purchased or rated [Linden et al. 2003; Ekstrand et al. 2011]. Item-based algorithms are then faster than User-based CF even for extremely large datasets.

## 2.3 Overview and limitations of Memory-based CF

Memory-based CF algorithms usually rely on neighborhood discovery, and are defined by using the user-item matrix interactions (ratings, likes, ...). Computations are made over the entire interaction matrix (for User-based CF) each time a prediction or a recommendation is needed. Despite their success, Memory-based CF algorithms have two major limitations: the first is related to sparsity and the second is related to scalability [Karypis 2001].

In general, users of commercial platforms do not interact with the entire set of items. They usually interact with a very small portion of the items' set (the amount of historical information for each user and for each item is often quite limited). In particular, since the correlation coefficient (similarity score) is only defined between users who have rated at least two products in common, many pairs of users have no correlation at all [Billsus & Pazzani 1998; Karypis 2001]. This problem is known as *reduced coverage* [Sarwar et al. 2000], and is due to sparse ratings of users. Accordingly, a recommender system based on nearest neighbor algorithms may be unable to make any item recommendations for a particular user. As a result, the accuracy of recommendations may be poor [Karypis 2001; Sarwar et al. 2001]. Moreover, with millions of users and items, User-based recommender systems suffer serious scalability problems that affect real-time performance : The computational complexity of these methods grows linearly with the number of customers. Table 3 presents an overview of Memory-based CF algorithms.

*Table 3: Overview of Memory-based CF algorithms*

| **Representative techniques** | **Main advantages** | **Main shortcomings** |
|---|---|---|
| **User-based (user-to-user) CF** [Resnick et al. 1994]:<br>• Find k users similar to an active user u,<br>• Find and recommend the top N items purchased by these k users and not purchased by user u. | • **Simplicity** : they are easy to implement; no costly training phases,<br>• **Explainability** : recommendations provided to a user can easily be justify,<br>• Need not to consider the content of items or users. | • **Scalability** (for User-based): computations grow linearly with the number of users and items,<br>• **Cold start** problem : unable to make recommendations for new items or new users,<br>• Depends on interactions between users and items,<br>• **Data sparsity** (not enough users/items interactions): users interact only with a tiny set of items. |
| **Item-based (item-to-item) CF** [Karypis 2001; Sarwar et al. 2001; Linden et al. 2003]:<br>• Find the list of similar items of an item i, not already purchased by an active user u and similar to items already purchased by u,<br>• Sort the result in decreasing order to be the top-N recommendation. | In addition to advantages of User-based CF:<br>• **Stability** : little affected by the constant addition of new information,<br>• **Scalability** : The use of pre-computed item-to-item similarities can significantly improve the real-time recommendation complexity. | |

# 3 Model-Based Collaborative Filtering

Model-based algorithms have been designed to address the limitations of Memory-based CF. Instead of computing recommendations on the entire user-item interactions database, a model-based approach builds a generic model from these interactions, which can be used each time a recommendation is needed [Breese et al. 1998; Parhi et al. 2017]. This allows the system to recognize complex patterns from the training data, and then make more intuitive predictions for new data based on the learned model [Su & Khoshgoftaar 2009]. Herein, every new information from any user outdates the model. The model then has to be rebuilt for any new information [Bobadilla et al. 2013]. Several varieties of models have been investigated, such as D*imensionality reduction* models [Billsus & Pazzani 1998; Koren et al. 2009], *Clustering* models [Xu & Tian 2015; Saxena et al. 2017], *Regression* models [Vucetic & Obradovic 2005; Hu & Li 2018; Steck 2019], *Neural Networks* [Feigl & Bogdan 2017; He et al. 2017a; He,Liao,Zhang,Nie,Hu & Chua 2017b; Zhang et al. 2018] and *Deep Learning* models [Mu 2018; Zhang,Yao,Sun & Tay 2019]. Since 2009, due to the success of Matrix Factorization [Koren et al. 2009] on the Netflix Price[3], a particular attention has been made on dimensionality reduction techniques. For this reason, in this section, we are focusing our study on dimensionality reduction based collaborative filtering.

To address the problem of high-level sparsity, several *Dimensionality Reduction* techniques have been used such as *Latent Semantic Indexing* (LSI) [Deerwester,Dumais,Furnas,Landauer & Harshman 1990], *Matrix Factorization models* [Koren et al. 2009] and its variants like *Singular Value Decomposition* (SVD) [Billsus & Pazzani 1998; Sarwar et al. 2000], *Regularized* SVD [Paterek 2007], *Non-negative Matrix Factorization* (NMF) [Wang & Zhang 2013; Hernando et al. 2016], *Probabilistic Matrix Factorization* (PMF) [Mnih & Salakhutdinov 2008], *Bayesian Matrix Factorization* (BMF) [Lakshminarayanan et al. 2011], Explainable Matrix Factorization (EMF) [Abdollahi & Nasraoui 2016; Abdollahi & Nasraoui 2017; Wang et al. 2018].

## 3.1 Singular Value Decomposition (SVD)

Since nearest neighbors-based recommendations do not deal with large and sparse data, the idea behind matrix factorization models is to reduce the degree of sparsity by reducing dimension of the data. SVD is a well-known matrix factorization model that captures latent relationships between users and items and produces a low-dimensional representation of the original user-item space that allows the computation of neighborhoods in the reduced space [Sarwar et al. 2000; Ekstrand et al. 2011]. SVD factors the $m \times n$ ratings matrix $R$ as a product of three matrices :

$$R = P\Sigma Q^\top, \qquad (8)$$

where $P$, $Q$ are two orthogonal matrices of size $m \times \hat{k}$ and $n \times \hat{k}$ respectively (where $\hat{k}$ is the rank of $R$) and $\Sigma$ a diagonal matrix of size $\hat{k} \times \hat{k}$ having all singular values of the rating matrix $R$ as its diagonal entries [Billsus & Pazzani 1998; Sarwar et al. 2000; Ekstrand et al. 2011; Bokde,Girase & Mukhopadhyay 2015]. It is possible to truncate the $\hat{k} \times \hat{k}$ matrix $\Sigma$ by only retaining its $k$ largest singular values to yield $\Sigma_k$, with $k < \hat{k}$. By reducing matrices $P$ and $Q$ accordingly, the reconstructed matrix $R_k = P_k \Sigma_k Q_k^\top$ will be the closest rank-$k$ matrix to $R$ (see Figure 1.a) [Sarwar et al. 2000; Ekstrand et al. 2011].

---

3  The **Netflix Price** (https://netflixprize.com) sought to substantially improve the accuracy of predictions about how much someone is going to enjoy a movie based on their movie preferences.

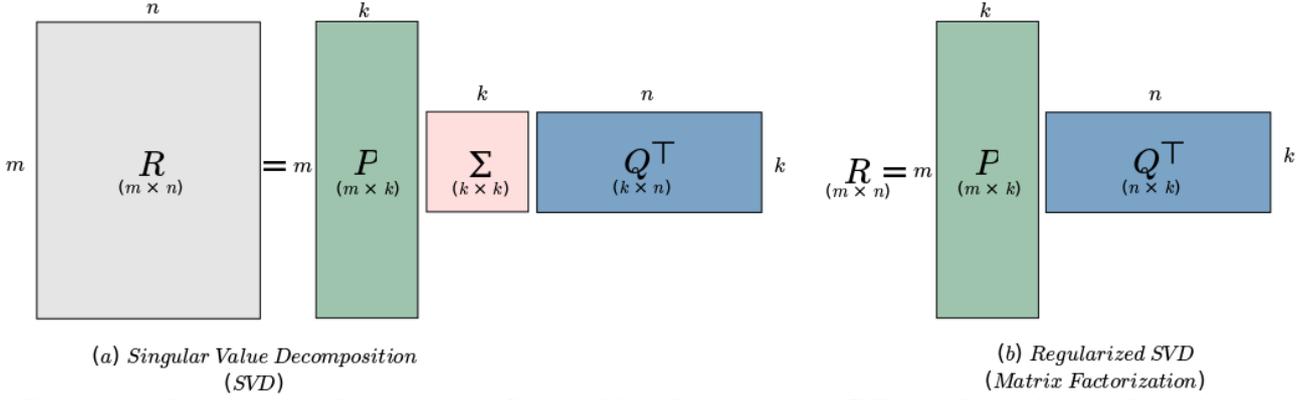

*Figure 1: (a) Dimensionality Reduction with Singular Value Decomposition (SVD). (b) Dimentionality Reduction with Matrix Factorization (MF). [Sarwar et al. 2000; Ekstrand et al. 2011].*

With the truncated SVD recommender system, user's and item's latent spaces are defined respectively by $P_k \cdot \sqrt{\Sigma_k}$ and $\sqrt{\Sigma_k} \cdot Q_k^\top$. In other words, the $u^{th}$ row of $P_k \cdot \sqrt{\Sigma_k}$ (resp. the $i^{th}$ column of $\sqrt{\Sigma_k} \cdot Q_k^\top$) represents the latent factors of user $u$ (resp. the latent factors of item $i$). The reduced latent space can then be used to perform rating-predictions as well as top-N recommendations for users: given user $u$ and item $i$, the predicted rating of user $u$ on item $i$, denoted by $\hat{R}_{u,i}$ is computed with equation 9:

$$\hat{R}_{u,i} = \left[P_k \cdot \sqrt{\Sigma_k}\right]_u \left[\sqrt{\Sigma_k} \cdot Q_k^\top\right]_i. \tag{9}$$

Applying SVD for CF recommendation often raises difficulties due to the high portion of missing values caused by sparseness in the user-item ratings matrix. Therefore, to factor the rating matrix, missing values must be filled with reasonable default values (this method is called imputation [Su et al. 2008] ). Sarwar et al. [Sarwar et al. 2000] found the item's mean rating to be a useful default value. Adding ratings normalization by subtracting the user mean rating or other baseline predictor can improve accuracy. The SVD based recommendation algorithm consists of the following steps :

Given the rating matrix $R$,

1. Find the normalized rating-matrix $R_{norm}$,
2. Factor $R_{norm}$ to obtain matrices $P$, $\Sigma$ and $Q$,
3. Reduce $\Sigma$ to dimension $k$ to obtain $\Sigma_k$,
4. Compute the square-root of $\Sigma_k$ to obtain $\sqrt{\Sigma_k}$,
5. Compute $P_k \cdot \sqrt{\Sigma_k}$ and $\sqrt{\Sigma_k} \cdot Q_k^\top$ that will be used to compute recommendation scores for any user and items.

An iterative approach of SVD called Regularized SVD, commonly known as Matrix Factorization (Section 3.2) [Koren et al. 2009] uses the gradient-descent method to estimate the resulting matrices. The obtained model will not be a true SVD of the rating-matrix, as the component matrices are no longer orthogonal, but tends to be more accurate at predicting unseen preferences than the standard SVD [Ekstrand et al. 2011].

## 3.2  Matrix Factorization (MF)

The idea behind matrix factorization also known as *Regularized SVD* [Paterek 2007; Zheng,Ding & Nie 2018], is the same as that of the standard SVD : represent users and items in a lower dimensional latent space. The original algorithm[4] factorized the user-item rating-matrix as the product of two lower dimensional matrices $P$ and $Q$, representing *latent factors* of users and items respectively (see Figure 1.b). If $k$ defines the number of latent factors for users and items, the dimension of $P$ and $Q$ are respectively $m \times k$ and $n \times k$. More specifically, each user $u$ is associated with a vector $P_u \in \mathbb{R}^k$ and each item $i$ with $Q_i \in \mathbb{R}^k$. $Q_i$ measure the extent to which item $i$ possesses those factors and $P_u$ measure the extent of interest user $u$ has in items that have high values on the corresponding factors [Koren et al. 2009; Bokde et al. 2015]. Therefore, the dot product of $Q_i$ and $P_u$ approximates the user $u$'s rating on item $i$ with $\hat{R}_{u,i}$ defined by equation 10:

$$\hat{R}_{ui} = Q_i^\top P_u. \tag{10}$$

The major challenge lies in computing the mapping of each item and user to factor vectors $Q_i, P_u \in \mathbb{R}^k$. After the recommender system completes this mapping, it can easily estimate the rating a user will give to any item [Koren et al. 2009; Parhi et al. 2017]. In the training stage, the model includes a regularizer to avoid overfitting. Therefore, the system minimizes the regularized squared error on the set of known ratings. The cost function $J$ is defined by equation 11:

$$J_{(P,Q)} = \frac{1}{2} \sum_{(u,i) \in \kappa} (R_{u,i} - \hat{R}_{u,i})^2 - \frac{\lambda}{2}(||P_u||^2 + ||Q_i||^2), \tag{11}$$

where $\kappa$ is the set of the $(u,i)$ pairs for which $R_{u,i}$ is known (the training set) [Paterek 2007; Koren et al. 2009; Bokde et al. 2015; Zheng et al. 2018]. The parameter $\lambda$ controls the extent of regularization and is usually determined by cross-validation.

To minimize the cost function $J$, the matrix factorization algorithm computes $\hat{R}_{u,i}$ and the associated error $e_{u,i}$ for each training case with the Mean Absolute Error (MAE) metric [Koren et al. 2009]. Parameters $P_u$ and $Q_i$ are then updated using the stochastic gradient descent algorithm with the following update rules (equations 12 and 13) :

$$Q_i \leftarrow Q_i + \alpha \cdot (e_{u,i} \cdot P_u - \lambda \cdot Q_i), \tag{12}$$
$$P_u \leftarrow P_u + \alpha \cdot (e_{u,i} \cdot Q_i - \lambda \cdot P_u), \tag{13}$$

where $\alpha$ is the learning rate. The following algorithm summarizes the matrix factorization model :

1. Initialize matrices $P$ and $Q$ with random values,
2. For each training example $R_{u,i}$, with $(u,i) \in \kappa$ :
   (a) Compute $\hat{R}_{u,i} = Q_i^\top P_u$,
   (b) Compute the prediction error $e_{u,i} = |R_{u,i} - \hat{R}_{u,i}|$,
   (c) Update parameters $Q_i$ and $P_u$ with Stochastic Gradient Descent with equations 12 and 13 respectively.
3. Repeat step 2 until obtaining satisfactory parameters (ideally at the global minimum).

A probabilistic foundation of Matrix Factorization (*Probabilistic Matrix Factorization* - PMF) is defined in [Mnih & Salakhutdinov 2008].

## 3.3 Probabilistic Matrix Factorization (PMF)

Probabilistic algorithms aim to build probabilistic models for user behavior and use those models to predict future behaviors. The core idea of these methods is to compute either the probability $\Pr(i|u)$ that user $u$ will purchase item $i$, or the probability $\Pr(r_{u,i}|u)$ that user $u$ will rate item $i$ [Hofmann 1999; Ekstrand et al. 2011]. These algorithms generally decompose the probability $\Pr(i|u)$ by introducing a set $Z$ of latent factors and then, decomposes $\Pr(i|u)$ as follows (equation 14)

$$\Pr(i|u) = \sum_{z \in Z} \Pr(i|z) \Pr(z|u), \qquad (14)$$

where $z$ is a latent factor (or feature), $\Pr(z|u)$ the probability that user $u$ picks a random feature $z$ and $\Pr(i|z)$ the probability to pick a random item $i$ given a feature $z$. As illustrated in Figure 2.a and detailed in [Hofmann 1999], a user selects an item by picking the associated factor $z$. This model is known as the *Probabilistic Latent Semantic Analysis* (PLSA). The PLSA model has the same form as SVD [Sarwar et al. 2000] or MF [Koren et al. 2009] ($R = P^\top Q$) except that $P$ and $Q$ are stochastic and not orthogonal. One downside of PLSA is that it's pretty slow to learn. An alternative probabilistic model is Probabilistic Matrix Factorization (PMF) [Mnih & Salakhutdinov 2008]. With PMF, ratings are drawn from the normal distributions.

---

[4] Funk, S (2006). Netflix Update: Try This at Home. https://sifter.org/~simon/journal/20061211.html

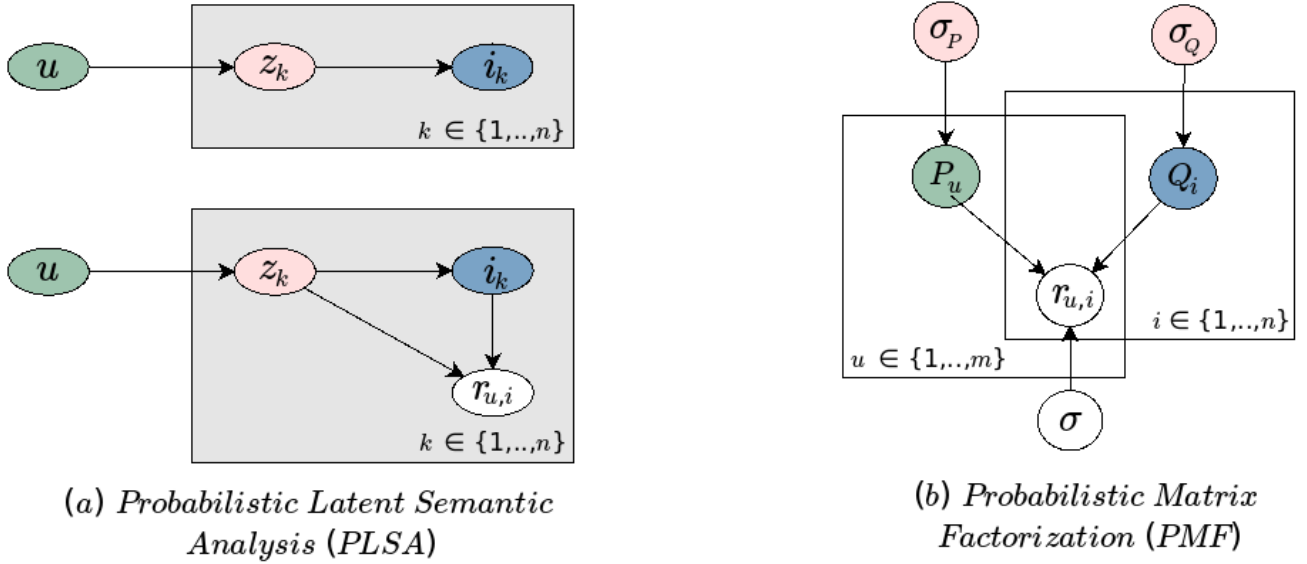

*Figure 2: Dimensionality Reduction : Probabilistic Models [Mnih & Salakhutdinov 2008; Ekstrand et al. 2011].*

The conditional distribution over the observed ratings $R$ (the likelihood term) and the prior distributions over the latent variables $P$ and $Q$ are given by equations 15, 16 and 17, where the mean for $R_{u,i}$ is denoted by $Q_i^\top P_u$ (see Figure 2.b).

$$\Pr(R|P, Q, \sigma^2) = \prod_{u=1}^{m} \prod_{i=1}^{n} [\mathcal{N}(R_{u,i}|P_u^\top Q_i, \sigma^2)]^{I_{u,i}}, \qquad (15)$$

$$\Pr(P|\sigma_P^2) = \prod_{u=1}^{m} \mathcal{N}(P_u|0, \sigma_P^2 I), \qquad (16)$$

$$\Pr(Q|\sigma_Q^2) = \prod_{i=1}^{n} \mathcal{N}(Q_i|0, \sigma_Q^2 I), \qquad (17)$$

where $\mathcal{N}(x|\mu, \sigma^2)$ is the probability density function of the Gaussian distribution with mean $\mu$ and variance $\sigma^2$, and $I_{u,i}$ is the indicator function that is equals to 1 if user $u$ rated item $i$ and equals to zero otherwise. The complexity of the model is controlled by placing a zero-mean spherical Gaussian priors on $P$ and $Q$, ensuring that latent variables do not grow too far from zero.

The learning procedure of the Regularized SVD (Matrix Factorization) can be applied to PMF by extending the number of regularization parameters. The goal is to minimize the sum-of-squared-error objective function with quadratic regularization terms (equation 18):

$$J_{(P,Q)} = \frac{1}{2} \sum_{(u,i) \in \kappa} (R_{u,i} - \hat{R}_{u,i})^2 - \frac{\lambda_P}{2} \sum_u ||P_u||^2_{Frob} + \frac{\lambda_Q}{2} \sum_i ||Q_i||^2_{Frob}, \qquad (18)$$

where $\lambda_P = \sigma^2/\sigma_P^2$, $\lambda_Q = \sigma^2/\sigma_Q^2$ are the regularization parameters and $||\cdot||^2_{Frob}$ is Frobenius norm. PMF is faster to train than PLSA and showed promising results compared to SVD as experimented in [Mnih & Salakhutdinov 2008]. Despite the effectiveness of PMF, it also has limitations. The basic assumption of PMF is that both users and items are completely independent, while in the field of recommendation, there may be some correlation between different features [Zhang et al. 2018].

## 3.4 Non-negative Matrix Factorization (NMF)

Component vectors of $P$ and $Q$ generated either by MF [Paterek 2007; Koren et al. 2009] or PMF [Mnih & Salakhutdinov 2008] are hard to understand, since their components can take arbitrary positive and/or negative values. They do not have any straightforward probabilistic interpretation. Therefore, as illustrated in [Hernando et al. 2016] these models cannot justify the prediction they make since the components of $P_u$ and $Q_i$ are difficult to interpret. With NMF (Non-negative MF) [Lee & Seung 1999; Pauca,Piper & Plemmons 2006; Cai,He,Wu & Han 2008; Wang & Zhang 2013; Gillis 2014; Hernando et al. 2016], a particular case of MF where the lower-rank matrices $P$ and $Q$ are non-negative matrices ($P \in \mathbb{R}_{\geq 0}^{m \times k}$ and $Q \in \mathbb{R}_{\geq 0}^{m \times k}$), it is possible to tackle this issue. In NMF [Lee & Seung 1999; Cai et al. 2008; Hernando et al. 2016], latent factors $P_{u,l}, Q_{i,l} \in [0,1]$ with $l \in \{1,..,k\}$. This allows a probabilistic interpretation : (a) latent factors represent groups of users who share the same interests, (b) the value of $P_{u,l}$ represents the probability that user $u$ belongs to the group $l$ of users and (c) the value $Q_{i,l}$ represents the probability that users in the group $l$ like item $i$. This property allows the justification and understanding of all recommendations provided by the model [Hernando et al. 2016]. The objective function of NMF is the same as that of the PMF model (see equation 18). However, to ensure that the values of $P$ and $Q$ are kept between 0 and 1, it is necessary to use the multiplicative update rule (see equations 19 and 20) [Lee & Seung 1999; Lee & Seung 2001] instead of the additive update rule (equations 12 and 13).

$$P_{u,l} \leftarrow P_{u,l} \cdot \frac{\sum_{i \in I_u} Q_{i,l} \cdot R_{u,i}}{\sum_{i \in I_u} Q_{i,l} \cdot \hat{R}_{u,i} + \lambda_P |I_u| P_{u,l}}, \qquad (19)$$

$$Q_{i,l} \leftarrow Q_{i,l} \cdot \frac{\sum_{u \in U_i} P_{u,l} \cdot R_{u,i}}{\sum_{u \in U_i} P_{u,l} \cdot \hat{R}_{u,i} + \lambda_Q |U_i| Q_{i,l}}, \qquad (20)$$

where $I_u$ is the set of items purchased/rated by user $u$ and $U_i$ is the set of users who purchased/rated item $i$. $\lambda_P$ and $\lambda_Q$ are the regularizer parameters.

NMF models are also known to be explainable models [Cai et al. 2008; Wang & Zhang 2013; Hernando et al. 2016]. Indeed, since classical MF models cannot justify their recommendations, justifications or explanations can be added as input to the model in order to provide explainable recommendations.

## 3.5 Explainable Matrix Factorization (EMF)

In [Abdollahi & Nasraoui 2017], Abdollahi and Nasraoui proposed an Explainable Matrix Factorization (EMF) model. The proposed model is a probabilistic formulation of their previous EMF model [Abdollahi & Nasraoui 2016], where the explanation is based on the neighborhood. The neighborhood is either User-based or Item-based (Section 2.2.2). The explainability score (for User-based neighbors style explanation) of the rating of user $u$ on item $i$, denoted by $Expl_{u,i}$ is computed with equation 21 [Abdollahi & Nasraoui 2017]:

$$Expl_{u,i} = E(R_{v,i}|N_u) = \sum_{x \in X} x \cdot Pr(R_{v,i} = x|v \in N_u), \tag{21}$$

where

$$Pr(R_{v,i} = x|v \in N_u) = \frac{|N_u \cap U_{i,x}|}{|N_u|} \tag{22}$$

$U_{i,x}$ is the set of users who have given the same rating $x$ to item $i$ and $N_u$ is the set of $k$-nearest neighbors for user $u$. As you may notice, $E(x|N)$ is the expectation of rating $x$ given a set of $N$ users. Similarly, for Item-based neighbor style explanation, the explanation score is given by equation 23.

$$Expl_{u,j} = E(R_{u,j}|N_i) = \sum_{x \in X} x \cdot Pr(R_{u,j} = x|j \in N_j). \tag{23}$$

Finally, the explanation weight $W_{u,i}$ of the rating given by user $u$ on item $i$ is

$$W_{ui} = \begin{cases} Expl_{u,i} \; if \; Expl_{u,i} > \theta \\ 0 \; otherwise \end{cases}, \tag{24}$$

where $\theta$ is an optimal threshold above which item $i$ is considered to be explainable for user $u$. $W_{u,i}$ is therefore the extent to which item $i$ is explainable for user $u$. The new objective function [Abdollahi & Nasraoui 2017] can be formulated as in equation 25:

$$\min_{P,Q} \sum_{(u,i) \in \kappa} (R_{u,i} - \hat{R}_{u,i})^2 + \frac{\beta}{2}(||P_u||^2 + ||Q_i||^2) + \frac{\lambda}{2}(P_u - Q_i)^2 W_{ui}, \tag{25}$$

where $\beta$ is the coefficient of the L2 regularization term and $\lambda$ is an explainability regularization coefficient. Similarity-Based EMF (SEMF) and Neighborhood-based EMF [Wang et al. 2018] are extensions of EMF. In [Wang et al. 2018] the expected value of the explainability score is obtained more reasonably through the Pearson Correlation Coefficient of users or items. SEMF is based on the assumption that if users have a close relationship, the influence of neighbors on users should be larger. Then the high rating items of users who are similar to an active user are also more explainable. Therefore, instead of using the empirical conditional probability of rating $W_{u,i}$ as explanation, SEMF uses the Pearson Correlation Coefficient $W_s(u,i)$ [Resnick et al. 1994]. SEMF can be extended over the entire users or items to produce User-based Neighborhood-based EMF or Item-based Neighborhood-based EMF [Wang et al. 2018]. Another extension of EMF is Novel and Explainable Matrix Factorization (NEMF) [Coba,Symeonidis & Zanker 2019]. This particular model includes novelty information in EMF in order to provide novel and explainable recommendations to users.

## 3.6 Overview of dimensionality reduction techniques

Despite the effectiveness of MF models in improving traditional memory-based CF, they are three general issues that arise with this family of models [Shenbin,Alekseev,Tutubalina,Malykh &

Nikolenko 2020]. First, the number of parameters in MF models is huge ($m \times k + n \times k$) : it linearly depends on both the number of users and items, which slows down model learning and may lead to overfitting. Second, making predictions for new users/items based on their ratings implies running optimization procedures in order to find the corresponding user/item embedding. Third, high data sparsity could also lead to overfitting. Moreover, MF models use linear transformations to learn these latent features [Zhang et al. 2019; Shenbin et al. 2020]. However, linear models are not complex enough to capture nonlinear patterns from user-item interactions. The learned features are then not sufficiently effective especially when the rating matrix is highly sparse. Table 4 presents an overview of matrix factorization algorithms.

*Table 4: Overview of Dimensionality Reduction techniques*

| **Models** | **Rules** |
|---|---|
| Singular Value Decomposition (SVD) [Billsus & Pazzani 1998; Sarwar et al. 2000] | Decomposes the rating matrix $R$ into three low dimensional matrices $P$, $\Sigma$ and $Q$ of dimensions $m \times \hat{k}$, $\hat{k} \times \hat{k}$ and $n \times \hat{k}$ respectively, where $\hat{k}$ is the rank of the rating matrix, then takes the $k$ largest singular values of $\Sigma$ to yield $\Sigma_k$ and reduces $P$ and $Q$ accordingly to obtain $P_k$ and $Q_k$. SVD cannot be applied on a matrix that contains missing values. |
| Regularized SVD (Matrix Factorization) [Paterek 2007; Koren et al. 2009] | Decomposes $R$ into two orthogonal matrices $P$ and $Q$. It is not necessary to fill missing values before applying Regularized SVD. The learning procedure can be done with gradient descent over all observed ratings. |
| Probabilistic Matrix Factorization (PMF) [Mnih & Salakhutdinov 2008] | Decomposes $R$ into two stochastic matrices $P$ and $Q$. It assumes that users and items are completely independent and ratings are drawn from the normal distributions. |
| Non-negative Matrix Factorization (NMF) [Lee & Seung 1999; Cai et al. 2008; Hernando et al. 2016] | Values of $P$ and $Q$ in the MF model are non-interpretable (they can be either positive or negative). NMF restrict values of $P$ and $Q$ between 0 and 1 that allows a probabilistic interpretation of latent factors. |
| Explainable Matrix Factorization (EMF) [Abdollahi & Nasraoui 2016; Abdollahi & Nasraoui 2017; Wang et al. 2018; Coba et al. 2019] | Addresses the explainability of MF in a different way compared to NMF. Here, an item $i$ is considered to be explainable for user $u$ if a considerable number of user $u$'s neighbors rated item $i$. |

# 4 Evaluation Metrics for CF algorithms

Before applying a recommender system in real life applications, it is important to measure its aptitude to capture the interest of a particular user in order to provide acceptable recommendations [Bobadilla,Hernando,Ortega & Bernal 2011]. The effectiveness of a recommendation system then depends on its ability to measure its own performance and fine-tune the system to do better on test and real data [Breese et al. 1998; Cacheda,Carneiro,Fernández & Formoso 2011]. However, evaluating recommender systems and their algorithms is inherently difficult for several reasons : first, different algorithms may be better or worse on different data sets; second, the goals for which an evaluation is performed may differ [Herlocker et al. 2004; Cacheda et al. 2011].

During the two last decades, dozens of metrics have been proposed in order to evaluate the behavior of recommender systems, from how accurate the system is to how satisfactory recommendations could be. Evaluation metrics can be categorized in four main groups [Herlocker et al. 2004; Cacheda et al. 2011; Bobadilla et al. 2013; Mahdi Seyednezhad,Cozart,Bowllan & Smith 2018] : (a) prediction metrics, such as accuracy (b) set of recommendation metrics, such as precision and recall (c) rank of recommendations metrics like half-life and (d) diversity and novelty.

## 4.1 Prediction accuracy

*Prediction accuracy* measures the difference between the rating the system predicts and the real rating [Cacheda et al. 2011]. This evaluation metric is most suitable for the task of rating prediction. Assuming that observed ratings are splitted into train and test sets, prediction accuracy evaluates the ability of the system to behave accurately on the test set. The most popular of these kinds of metrics are *Mean Absolute Error (MAE), Mean Squared Error (MSE), Root Mean Squared Error (RMSE)* and *Normalized Mean Absolute Error (NMAE)* [Herlocker et al. 2004; Willmott & Matsuura 2005; Bobadilla et al. 2011; Bobadilla et al. 2013].

Let us define $U$ as the set of users in the recommender system, $I$ as the set of items, $R_{u,i}$ the rating of user $u$ on item $i$, $\emptyset$ the lack of rating ($R_{u,i} = \emptyset$ means user $u$ has not rated item $i$), $\hat{R}_{u,i}$ the prediction of user $u$ on item $i$.

If $O_u = \{i \in I | \hat{R}_{u,i} \neq \emptyset \wedge R_{u,i} \neq \emptyset\}$ is the set of items rated by user $u$ and having prediction values, we can define the *MAE* (equation 26) and *RMSE* (equation 27) of the system as the average of the user's MAE and RMSE [Bobadilla et al. 2011; Bobadilla et al. 2013]. The prediction's error is then defined by the absolute difference between prediction and real value $|\hat{R}_{u,i} - R_{u,i}|$.

$$MAE = \frac{1}{\#U} \sum_{u \in U} (\frac{1}{\#O_u} \sum_{i \in O_u} |\hat{R}_{u,i} - R_{u,i}|), \qquad (26)$$

$$RMSE = \frac{1}{\#U} \sum_{u \in U} \sqrt{\frac{1}{\#O_u} \sum_{i \in O_u} (\hat{R}_{u,i} - R_{u,i})^2}. \qquad (27)$$

Despite their limitations when evaluating systems focused on recommending a certain number of items, the simplicity of their calculations and statistical properties have made MAE and RMSE metrics the most popular when evaluating rating predictions for recommendation systems [Herlocker et al. 2004; Cacheda et al. 2011].

Another important and commonly used metric is *coverage*, which measures the percentage of items for which the system is able to provide prediction for [Herlocker et al. 2004]. The coverage of an algorithm is especially important when the system must find all the good items and not be limited to recommending just a few. According to [Bobadilla et al. 2011; Bobadilla et al. 2013], coverage represents the percentage of situations in which at least one k-neighbor of each active user can rate an item that has not been rated yet by that active user. If $K_{u,i}$ defines the set of neighbors of user $u$ who have rated item $i$, then the coverage $c_u$ of user $u$ is defined by equation 28 :

$$c_u = 100 \times \frac{\#C_u}{\#D_u}, \text{ with } \begin{cases} C_u = \{i \in I | R_{u,i} = \emptyset \land K_{u,i} \neq \varnothing\} \\ \text{and} \\ D_u = \{i \in I | R_{u,i} = \emptyset\} \end{cases}, \qquad (28)$$

where $C_u$ is the set of items not rated by user $u$ but rated by at least one of his $k$-neighbors and $D_u$ the set of items not rated user $u$. The total coverage (equation 29) is then defined as the mean of all $c_u$ by :

$$coverage = \frac{1}{\#U} \sum_{u \in U} c_u. \qquad (29)$$

A high prediction accuracy in a recommender system doesn't guarantee the system will provide good recommendations. The confidence of users (or his satisfaction) for a certain recommender system does not depend directly on the accuracy for the set of possible predictions. A user rather gains confidence on the recommender system when he agrees with a reduced set of recommendations made by the system [Bobadilla et al. 2011; Bobadilla et al. 2013].

Rating predictions are made over items that users have already seen, but the more important thing to know is how well the system will behave on items they have neither seen nor purchased before. Moreover, prediction accuracy is suitable when a large number of observed ratings is available, ratings without which training and validation cannot be done effectively. When only a list of items rated by a user is known, it will be preferable to combine prediction accuracy with other evaluation metrics such as Quality of recommendations.

## 4.2 Quality of set of recommendations

*Quality of set of recommendations* is appropriate when a small number of observed ratings is known, and it is used to evaluate the relevance of a set of recommended items [Desrosiers & Karypis 2011] . The quality of set of recommendations is usually computed by the following three most popular metrics : (1) *Precision*, which indicates the proportion of relevant recommended items from the total number of recommended items, (2) *Recall*, which indicates the proportion of relevant recommended items from the number of relevant items, and (3) *F1-score*, which is a combination of *Precision* and *Recall* [Sarwar et al. 2000; Herlocker et al. 2004; Bobadilla et al. 2011; Bobadilla et al. 2013; Mahdi Seyednezhad et al. 2018].

Let consider $Z_u$, the set of $n$ recommended items to user $u$, $r_u \in Z_u$ the set of relevant items recommended to $u$ and its opposite $r_u^c$ the set of relevant items not recommended to $u$. We consider that an item $i$ is relevant for recommendation to user $u$ if $R_{u,i} \geq \theta$, where $\theta$ is a threshold.

Assuming that all users accept $n$ test recommendations, the evaluations of precision (equation 30), recall (equation 31) and F1-score (equation 32) are obtained by making $n$ tests recommendations to the user u as follow :

$$precision = \frac{1}{\#U} \sum_{u \in U} \frac{\#r_u}{n}, \text{ with } r_u = \{i \in Z_u | R_{u,i} \geq \theta\} \qquad (30)$$

$$recall = \frac{1}{\#U} \sum_{u \in U} \frac{\#r_u}{\#r_u + \#r_u^c}, \text{ with } r_u^c = \{i \notin Z_u | R_{u,i} \geq \theta\} \qquad (31)$$

$$F1 = \frac{2 \times precision \times recall}{precision + recall} \qquad (32)$$

One important drawback of this evaluation metric is that all items in a recommendation list are considered equally interesting to a particular user [Desrosiers & Karypis 2011]. However, items in a list of recommendations should be ranked according to the preference of the user. The rank of recommended items is evaluated with the quality of list of recommendations.

## 4.3 Quality of list of recommendations

Recommender systems usually produce lists of items to users. These items are ranked according to the relevancy since users give more attention to the first items recommended in the list. Errors made on these items are then more serious than those of the last items in the list. The ranking measures consider this situation. The most used metrics for ranking measures are (1) *Mean Average Precision (MAP)* [Caragea et al. 2009] which is just the mean of the average precision (equation 30) over all users, (2) *Half-life* (equation 33) [Breese et al. 1998], which assumes an exponential decrease in the interest of users as they move away from the recommendations at the top and (3) D*iscounted Cumulative Gain* (equation 34) [Baltrunas,Makcinskas & Ricci 2010], where the decay is logarithmic :

$$Hl = \frac{1}{\#U} \sum_{u \in U} \sum_{i=1}^{N} \frac{\max(R_{u,p_i} - d, 0)}{2^{(i-1)/(\alpha-1)}}, \qquad (33)$$

$$DCG = \frac{1}{\#U} \sum_{u \in U} (R_{u,p_1} + \sum_{i=2}^{k} \frac{R_{u,p_i}}{\log_2(i)}), \qquad (34)$$

where $p_1, ..., p_n$ represent the recommended list, $R_{u,p_i}$ the true rating of the user $u$ for item $p_i$, $k$ the rank of the evaluated item, $d$ the default rating, and $\alpha$ the number of the item in the list such that there is a 50% chance the user will review that item.

## 4.4 Novelty and Diversity

In many applications such as e-commerce, recommendation systems must recommend novel items to users, because companies want to sell their new items as well and keep their clients satisfied. Further, some users may want to explore a new type of item. Therefore, a metric to evaluate recommendation systems based on this criterion would be useful. In this case, we must evaluate the extent to which a recommendation system can produce diverse items. Novelty and diversity are two main metrics that provide such evaluations [Zhang 2013; Mahdi Seyednezhad et al. 2018].

The novelty metric evaluates the degree of difference between the items recommended to and known by the user. The novelty of an item $i$ (equation 35) is measured according to three major criteria : (a) Unknown : $i$ is unknown to the user, (b) Satisfactory : $i$ is of interest for the user, (c) Dissimilarity : $i$ is dissimilar to items in the user's profile [Zhang 2013].

$$novelty_i = \frac{1}{\#Z_u - 1} \sum_{j \in Z_u} (1 - sim(i,j)), i \in Z_u \qquad (35)$$

where $sim(i,j)$ indicates item to item memory-based CF similarity measures and $Z_u$ represents the set of $n$ recommendations to user $u$.

Novelty is central for recommender systems, because there are some items which most of the users do not buy frequently (like refrigerators). Thus, if a user buys one of them, most likely he or she will not buy it again in the near future. Then, the recommender system should not continue to recommend it to the user. However, if the user tries to buy them again, the system should learn that and include them in the set of recommended items [Zhang 2013; Mahdi Seyednezhad et al. 2018].

The diversity of a set of recommended items (for example $Z_u$), indicates the degree of differentiation among recommended items [Bobadilla et al. 2013]. It is computed by summing over the similarity between pairs of recommended items and normalizing it (equation 36) :

$$diversity_{Z_u} = \frac{1}{\#Z_u(\#Z_u - 1)} \sum_{i \in Z_u} \sum_{j \in Z_u, j \neq i} (1 - sim(i,j)) \qquad (36)$$

Evaluating the diversity and the novelty of a recommender system ensures that the system will not recommend the same items over and over.

## 4.5 Overview of evaluation metrics

The current main evaluation metrics (detailed previously) are summarized in Table 5.

*Table 5: Overview of recommendation system evaluation metrics*

| Evaluation Metrics | Measures | Observations |
|---|---|---|
| **Accuracy** : useful when a large range of ratings is available. Appropriate for the *rating prediction*'s task | • **Mean Absolute Error (MEA).**<br>• **Root Mean Square Error (RMSE).**<br>• **Coverage** : percentage of items that can be recommended. | • Evaluate performances of recommendations over items already purchased (or rated) by users. But we need to know how they will behave on items the user has not purchased yet,<br>• A recommender system can be highly accurate, but provide poor recommendations. |
| **Quality of set of recommendation**: users purchased or rated only a few items; Only a small range of ratings is available. Appropriate to *recommend good items* | • **Precision**: proportion of relevant recommended items from the total number of recommended items.<br>• **Recall** : proportion of relevant recommended items from the number of relevant items.<br>• **F1-score** : combination of *precision* and *recall*. | **Note**: an item i is relevant to a user $u$ if $R_{u,i} > \theta$. Where is a threshold<br>• Evaluate the relevancy of a set of recommendations provided to a set of users<br>• Items in the same recommendation list are considered equally interesting to the user. |
| **Quality of list of recommendation**: rank measure of a list of recommended items. Users give greater importance to the first items recommended to them.<br>This metric Evaluates the utility of a ranked list to the user. | • **Half-life** (Hl) : assume that the interest of a user on the list of recommendations decreases exponentially as we go deeper in the list.<br>• **Discounted Cumulative Gain** (DCG): the decay here is logarithmic. | • Ranking evaluation of the top-N recommendations provided to the user. |
| **Novelty and Diversity** | • **Novelty**: evaluate the ability of the system to recommend items not already purchased by users.<br>• **Diversity**: measure the degree of differentiation among recommended items. | • The goal of a recommender system is to suggest to a user a set of items that he has not purchased yet. It would be inappropriate to constantly recommend items that the user has already seen.<br>• Diversity is quite important because it ensures that the system won't recommend all the time the same type of items. |

# 5 Comparative experimentation

In this section, we compare performances of all the models previously presented (Section 2 and 3). We use the MovieLens[5] benchmark datasets and evaluate these recommendation algorithms with the MAE evaluation metric (Section 4).

## 5.1 MovieLens Datasets

The GroupLens Research team[6] provides datasets that consist of user ratings (from 1 to 5) on movies. They are various sizes of the MovieLens datasets :

- **ML-100K** : 100,000 ratings from 1000 users on 1700 movies, released in April 1998,

- **ML-1M**   : 1 million ratings from 6000 users on 4000 movies, released in February 2003,

- **ML latest** : at the moment we are writing this article, the latest version of the MovieLens dataset contains 27 million ratings applied to 58,000 movies by 280,000 users. The small version of this dataset (MovieLens latest Small) contains 100,000 ratings applied to 9723 movies by 610 users. The latest MovieLens data set changes over time with the same permalink. Therefore, experimental results on this dataset may also change because the distribution of the dataset may change as new versions are released. Therefore, it is not recommended to use the latest small MovieLens dataset for research results. For this reason, we will use only the ML-100K and the ML-1M datasets in our experiments.

In this study we present the experimental testing of the previous models on the ML-100K and the ML-1M datasets and we compare their performances using various evaluation metrics.

## 5.2 Experimental results and performances comparison

### 5.2.1 User-based vs Item-based Collaborative Filtering

Table 6 presents a performance comparison of User-based and Item-based CF. We use two different datasets and compare the results over the Euclidean and the Cosine-based similarities. The experimental errors of the User-based algorithm with the Cosine distance, **0.75** for ML-100k and **0.73** for ML-1M, is lower than those with the Euclidean distance, **0.81** for both ML-100k and ML-1M. Similarly, the MAEs of the Item-based algorithm with the Cosine distance, **0.51** for ML-100k and **0.42** for ML-1M, is lower than those with the Euclidean distance, **0.83** for ML-100k and **0.82** for ML-1M. Either on the ML-100k dataset or the ML-1M dataset, the Cosine Similarity allows the model to reduce the MAE. This is why the Cosine similarity is generally preferred as a similarity measure compared to the Euclidean distance.

As stated in Section 2.2.2, Item-based collaborative filtering scales well for online recommendations and globally offer better performances compared to the User-based ones. As described is [Sarwar et al. 2001; Ekstrand et al. 2011], this is due to the fact that items similarities can be calculated in advance, allowing better scaling for online recommendations compared to the User-based algorithm.

---

5  **MovieLens**  (https://movielens.org/) is a web-based recommender system and virtual community that recommends movies to its users, based on their past preferences using collaborative filtering.

6  **GroupLens Research** (https://grouplens.org/) is a research laboratory from the University of Minnesota, that advances the theory and practice of social computing by building and understanding systems used by real people.

*Table 6: User-based vs Item-based performances (MAE) across MovieLens datasets with different similarity-based metrics.*

| Metric | Dataset | User-based | Item-based |
|---|---|---|---|
| **Euclidean distance** | ML-100k | 0.81 | 0.83 |
| | ML-1M | 0.81 | 0.82 |
| **Cosine distance** | ML-100k | 0.75 | 0.51 |
| | ML-1M | 0.73 | 0.42 |

### 5.2.2 Importance of ratings normalization

In this section, we demonstrate the importance of data normalization, when training recommendation models, by comparing the performances of Matrix Factorization models on raw vs normalized ratings. As we can observe in Table 7 that ratings normalization boosts the models' performances. Indeed, ratings normalization helps to reduce bias incorporated in data (see Section 3.1), hence the results.

*Table 7: MF vs NMF vs EMF performances (MAE) on MovieLens dataset trained on raw vs normalized ratings with k=10 on 10 epochs.*

| Preprocessing | Dataset | MF | NMF | EMF |
|---|---|---|---|---|
| Use raw ratings | ML-100k | 1.497 | 0.9510 | **0.797** |
| | ML-1M | 1.482 | 0.9567 | **0.76** |
| Normalized ratings | ML-100k | 0.828 | --- | **0.783** |
| | ML-1M | 0.825 | --- | **0.758** |

The MAEs of the MF model on the raw and normalized scores are **~1.48** and **~0.82**, respectively. Normalizing the scores led to an error reduction of **~45%** from the initial error (on the raw data). To measure the difference between these two values, we performed a t-test with 50 degrees of freedom and obtained a t-value of **3.48** and a p-value of **0.00077** which is far below the classical threshold of **0.05** below which the gap, between 1.48 and 0.82, can be considered significant. This simply means that we have less than a 77 in 100,000 chance of being wrong in stating that this difference is significant. We can therefore conclude that *normalizing the scores before training the MF model leads to a significant reduction of the MAE*. In the case of EMF, there are no great differences between the MAE errors on raw and normalized ratings (**~0.78 vs ~0.77**). This is due to the positive impact of *explainable scores* that allow the model to reduce bias in the raw ratings leading to better results. The NMF model cannot be trained on normalized ratings. Indeed, normalized ratings may contain negative values which are then not compatible with the *non-negativity* constraint of NMF. However, the experimental results in Table 7 show the improvement of NMF over MF.

To obtain results in Table 7, we trained MF, NMF and EMF with the same number of latent factors $k = 10$ and for 10 epochs for each dataset. Regarding the MAE on raw ratings, we can easily conclude that the EMF model offers better performances (on every dataset) compared to NMF which in turn is better than MF. The performance of EMF over MF and NMF is due to the introduction of explainable scores [Abdollahi & Nasraoui 2016; Abdollahi & Nasraoui 2017; Wang et al. 2018; Coba et al. 2019]. Similarly, the performance of NMF over MF is due to the introduction of *non-negativity* that allows a probabilistic interpretation of latent factors [Lee & Seung 1999; Cai et al. 2008; Wang & Zhang 2013; Hernando et al. 2016].

# 6   Conclusion

A full review of collaborative filtering recommender systems has been presented, as well as the most important evaluation metrics to consider when building such systems. Due to the explosion of online commerce, collaborative filtering has attracted increasing attention, and a variety of models from memory-based to model-based have been proposed over time. Given the current state of the art, we conclude that memory-based filtering (section 2), simple and easy to use algorithms, are efficient when user's ratings are not highly sparse [Resnick et al. 1994; Billsus & Pazzani 1998; Sarwar et al. 2001; Deshpande & Karypis 2004; Ekstrand et al. 2011]. Our review also established that User-based [Goldberg et al. 1992; Breese et al. 1998; Goldberg et al. 2001] and Item-based [Karypis 2001; Sarwar et al. 2001; Linden et al. 2003; Deshpande & Karypis 2004] collaborative filtering cannot easily handle the high sparsity of user-item interactions data. Moreover, in our experiments we have reviewed that memory-based CF systems suffer from scalability concerns due to online computations on huge amounts of data each time a recommendation is needed.

As we have described in section 3, a lot of model-based recommenders have been designed to address the limitations of their memory-based counterparts. Among the proposed approaches, not involving Neural Networks and Deep Learning, the most used are Dimensionality Reduction models. Several such methods have been covered [Lee & Seung 2001; Mnih & Salakhutdinov 2008; Koren et al. 2009; Wang & Zhang 2013; Abdollahi & Nasraoui 2016; Hernando et al. 2016]. In term of Dimensionality Reduction, Matrix Factorization (MF) [Sarwar et al. 2000; Mnih & Salakhutdinov 2008; Koren et al. 2009], Non-negative Matrix Factorization (NMF) [Lee & Seung 1999; Gillis 2014; Hernando et al. 2016] and Explainable Matrix Factorization (EMF) [Abdollahi & Nasraoui 2016; Abdollahi & Nasraoui 2017] models appear as the most important ones, due to their ability to capture user's preferences through latent factors. To complete further our analysis, our experimentation (Section 5) illustrates that normalizing the ratings leads to the improvement of the performances, **~45%** of improvement in the case of the MF model. The NMF model in turn does not support normalized ratings (due its non-negativity constraint), however, the probabilistic interpretation of its latent factors has reduced the error of MF, on raw ratings, by **~36%**. Finally, introducing explainable scores into the MF model, to obtain the EMF model, improved the error on the raw data by **~48%**, for the MF model, and **~18%** for the NFM model. As a result, we conclude that currently, EMF offers better performance than NMF which in turn is more efficient than MF. The source code of our experimentation is available on github (Resources section 7).

**Future works**

We have reviewed that in Explainable Matrix Factorization, the explainable scores are included in the matrix factorization (MF) model, which leads to better results (see Table 7) compared to MF and NMF. However, the latent factors in EMF may still be positive or negative values that are not interpretable. Since applying non-negativity to an MF model (hence the NMF model) yields higher performances of recommendations thanks to the interpretability of the latent factors, we form the following hypothesis : introducing non-negativity to EMF models will further improve the recommendation performances. Moreover, the resultant model will be a two-stage explainable model : (a) the probabilistic interpretation of latent factors and (b) the further explainability with explainable weights. Devising such a Non-negative Explainable Matrix Factorization (NEMF) is a research direction we are exploring.

Due to space limitation, we have not included the deep learning recommendation systems in this paper. A thorough study of these models is the subject of our future work.

# 7  Resources

All experiments have been made with Python 3.7 on the Jupyter notebook with Intel Core i5, 7[th] gen (2.5GHz) and 8Go RAM.

**Requirements**

- Matplotlib  == 3.2.2
- Numpy == 1.19.2
- Pandas == 1.0.5
- Scikit-learn  == 0.24.1
- Scikit-surprise == 1.1.1
- Scipy == 1.6.2

**Github repository**

Our experiments are available on the following github repository : https://github.com/nzhinusoftcm/review-on-collaborative-filtering/ . The repository contains the following notebooks :

- User-based Collaborative Filtering : implements the User-based recommendation system. https://github.com/nzhinusoftcm/review-on-collaborative-filtering/blob/master/2.User-basedCollaborativeFiltering.ipynb
- Item-based Collaborative Filtering : implements the Item-based recommendation system. https://github.com/nzhinusoftcm/review-on-collaborative-filtering/blob/master/3.Item-basedCollaborativeFiltering.ipynb

- Matrix Factorization : implements the Matrix Factorization Model. https://github.com/nzhinusoftcm/review-on-collaborative-filtering/blob/master/5.MatrixFactorization.ipynb

- Non-negative Matrix Factorization : implements the Non-negative Matrix Factorization model. https://github.com/nzhinusoftcm/review-on-collaborative-filtering/blob/master/6.NonNegativeMatrixFactorization.ipynb

- Explainable Matrix Factorization : implements the Explainable Matrix Factorization model. https://github.com/nzhinusoftcm/review-on-collaborative-filtering/blob/master/7.ExplainableMatrixFactorization.ipynb

- Performance measure. https://github.com/nzhinusoftcm/review-on-collaborative-filtering/blob/master/8.PerformancesMeasure.ipynb

**Notebooks with Google Collaboratory**

To run these notebooks with Google Collaboratory, we provide the following link.
https://colab.research.google.com/github/nzhinusoftcm/review-on-collaborative-filtering/